\documentclass[12pt]{article}

\usepackage{epsfig}
\usepackage{amsmath}
\usepackage{amssymb}

 \addtocounter{secnumdepth}{1}

\begin{document}

\title{Nuclear fission: The "onset of dissipation"\\
from a microscopic point of view}
\author{H. Hofmann $^{1}$, F.A. Ivanyuk$^{1,2}$, C. Rummel $^{1}$
and S. Yamaji$^3$\\ \small\it{1) Physik-Department der TU
M\"unchen, D-85747 Garching, Germany} \\ \small\it{2) Institute
for Nuclear Research, 03028 Kiev-28, Ukraine } \\ \small\it{3)
Cyclotron Lab., Riken, Wako-Shi, Saitama, 351-01 Japan }\\
\\Revised Version}
\date{August 01, 2001}
 \maketitle

\begin{abstract}
Semi-analytical expressions are suggested for the temperature
dependence of those combinations of transport coefficients which
govern the fission process. This is based on experience with
numerical calculations within the linear response approach and the
locally harmonic approximation. A reduced version of the latter is
seen to comply with Kramers' simplified picture of fission. It is
argued that for variable inertia his formula has to be
generalized, as already required by the need that for overdamped
motion the inertia must not appear at all. This situation may
already occur above $T\approx 2~\text{MeV}$, where the rate is
determined by the Smoluchowski equation. Consequently, comparison
with experimental results do not give information on the effective
damping rate, as often claimed, but on a special combination of
local stiffnesses and the friction coefficient calculated at the
barrier.

\vspace*{10pt}
 \noindent PACS number(s): 24.10.Pa, 24.75.+i,25.70.Jj,25.70.Gh
\end{abstract}

\newpage
 \tableofcontents

\newpage

\section{Introduction}\label{intro}

It is of considerable interest to understand the temperature
dependence of transport properties associated with slow collective
motion of large scale. Fission is a prime example, and indeed, for
this case there is growing experimental evidence \cite{pauthoe} -
\cite{strasb} that damping effectively increases with $T$. One
often tries to characterize this feature by one parameter, the
effective damping rate $\eta $ which is related to the equation of
average motion for a locally defined damped oscillator
\begin{equation}\label{diffeq} M \frac{d^2 q}{dt^2} +
\gamma \frac{dq}{dt} + Cq(t) = 0\ ,\end{equation} through
 \begin{equation}\label{defgameta}
     \eta  = \frac{\gamma}{2\sqrt{M\mid C \mid}} \ .
      \end{equation}
The $q=Q-Q_0$ measures the deviation of the collective variable
$Q$ from some fixed value $Q_0$. In the following we will also
need other combinations of inertia $M$, stiffness $C$ and friction
$\gamma$, namely
\begin{equation}\label{defgagaom}
\tau_{\text{coll}} = \frac{\gamma}{|C|}
=\frac{\hbar}{\Gamma_{\text{coll}}}\ ,\qquad \tau_{\text{kin}} =
\frac{M}{\gamma} = \frac{\hbar}{\Gamma_{\text{kin}}} \qquad
\text{and} \qquad \varpi ^2= {{\mid C \mid }\over M}\,.
      \end{equation}
The $\tau_{\text{coll}}$ sets the scale for (local) relaxation of
collective motion in a given potential of (local) stiffness $C$.
The $\tau_{\text{kin}}$, on the other hand, measures the
relaxation of the kinetic energy to the equilibrium value of the
Maxwell distribution. Typically, for slow collective motion we
expect this time to be smaller than the former. The limit of
overdamped motion applies for $\tau_{\text{kin}} \ll
\tau_{\text{coll}}$. Using the $\eta$ introduced in
(\ref{defgameta}) the following useful relation for their ratio is
easily verified
\begin{equation}\label{etaforms}2 \eta =
\frac{\Gamma_{\text{kin}}}{\hbar\varpi} =
\varpi\tau_{\text{coll}}\, =
\sqrt{\frac{\tau_{\text{coll}}}{\tau_{\text{kin}}}}\,.
\end{equation}
For a positive stiffness ($C>0$) and underdamped motion, the
$\varpi$ would be the frequency of the vibration and the
$\Gamma_{\text{kin}}$ its width. It should be noted that in the
literature a different notation is sometimes used where $\gamma$
stands for $\eta$, and often the $\Gamma_{\text{kin}}/\hbar$ is
referred to as $\beta$.

To understand the dynamics in phase space one also needs the
diffusion coefficients. At small temperatures they may deviate
from the classic Einstein relation (see \cite{hofrep}), but these
finer details will be neglected here. For such a situation  the
fission decay rate is commonly calculated within Kramers' "high
viscosity" limit \cite{kram}, for which the dependence on friction
is given by \begin{equation}\label{ratetdep}
\frac{R_K^{h.v.}(\eta_b)}{R_K^{h.v.}(\eta_b=0)}=\sqrt{1+\eta_b^2}
- \eta_b \equiv \left(\sqrt{1+\eta_b^2} + \eta_b\right)^{-1} \,.
\end{equation}
Here, the index "b" refers to the fact that the transport
coefficients are to be calculated at the barrier. It has been
reported, see e.g. Fig.5 of \cite{hobapa}, experimental data to
suggest, when analyzed on the basis of (\ref{ratetdep}), the
$\eta$ to be negligibly small at very low temperatures but to rise
more or less sharply around $T\simeq 1 ~\text{MeV}$. This result
is in qualitative agreement with microscopic calculations of the
transport coefficients within linear response theory \cite{hofrep,
hitrans, ivhopair}, although caution is still warranted here. Let
us leave aside the fact that at lower temperatures the one
dimensional potential may attain more structure than that found in
one minimum and one barrier, which is the picture underlying
formula (\ref{ratetdep}) and which for the sake of simplicity
shall be applied in the sequel. As will be demonstrated below,
even then it is not permissible to entirely parameterize the truly
complicated transport process by the single quantity $\eta$.
Rather, other combinations of $M,~\gamma$ and $C$  are needed for
more realistic descriptions. Moreover, one should be guided by the
theoretical fact, that not all transport coefficients are equally
well accessible both theoretically as well as numerically, as is
so for the inertia, for instance.

Unfortunately, when physicists address transport problems, all too
often one disregards the importance of the inertia, and in
particular its variation with the collective coordinate. Indeed,
in studies based on the Caldeira-Leggett Hamiltonian (see e.g.
\cite{hantalbor}) or in applications of the Random Matrix Theory
(see \cite{RMT-tran}) the inertia is the one of the unperturbed
collective part of the total Hamiltonian, treated as an (unknown)
parameter. In the case of nuclear physics the situation is more
complicated. Firstly, there, no unperturbed inertia exists at all;
it may only show up in the final effective equation of motion as
one manifestation of the existence of collective dynamics.
Secondly, the inertia $M$ may depend sensitively on the collective
degree of freedom. This feature is already well known from the
traditional case of undamped motion at zero thermal excitation
\cite{funnyhills}. There is no a priori reason why this should be
different at finite temperature, with perhaps two exceptions or
modifications. With increasing $T$ the $M$ gets close to the
liquid drop value \cite{hoyaje}, which only varies smoothly with
$Q$ and which is quite small. Simultaneously the friction strength
increases, so that one may quickly reach the situation of
overdamped motion, for which no trace of the inertia can be seen
anymore. Such features have been seen within the linear response
approach (see \cite{hofrep}), but to the best of our knowledge no
other transport model has so far addressed this question.

\section{Rate formulas} \label{rateform}

Like in the analysis underlying \cite{backetal, dioszegi} we want
to make use of a simple formula for the decay rate.  In slight
modification of Kramers' \cite{kram} classic one we
write\footnote{Since we are only looking at stationary situations
we leave out the time dependent factor which sometimes is taken
into account to simulate the "transient time" it takes before the
stationary current has built up.}:
\begin{equation}\label{krahiviin}
R_K^{h.v.}(\eta_b)=\frac{\varpi_a}{2\pi}\,\sqrt{\frac{M_a}{M_b}}
\exp(-E_b/T) \left(\sqrt{1+\eta_b^2} - \eta_b\right)\,.
\end{equation}
The indices "$a$" and "$b$" refer to the minimum and maximum of
the potential $V(Q)$, located at $Q_a$ and $Q_b$, respectively.
The $E_b$ stands for the height of the barrier $E_b=V(Q_b)
-V(Q_a)$. The factor $\sqrt{M_a/M_b}$, not contained in Kramers'
original work, is meant to account for the modification one gets
for variable inertia. Notice, that this inertia both influences
the current over the barrier as well as the number of "particles"
(phase space points) sitting in the well. Commonly both quantities
are calculated with the same $M$ which then drops out; see e.g.
eqs. (4.30) and (4.31) of \cite{hantalbor}. General reasons for
the presence of this additional factor will be given in Appendix
\ref{implvarin}. At first, in sect.\ref{decravarin} we follow the
more common derivation involving the construction of the densities
at the barrier and at the minimum, which in sect.\ref{smolueq} is
reduced to the Smoluchowsky limit (see below). In
sect.\ref{kram-stru} we follow the arguments of Strutinsky
\cite{strutkram} which lead to exactly the same formula.
Unfortunately, in \cite{strutkram} this feature is disguised by
the very fact that in the final expression eq.(16), like in large
parts of the derivation, primed and unprimed quantities are
interchanged. It is for this reason that we feel compelled to redo
the short calculation. The form (\ref{krahiviin}) has recently
been derived also in \cite{rumhof} by applying a generalized
version of the "Perturbed Static Path Approximation (PSPA)".

\begin{figure}
\begin{center}
\epsfig{figure=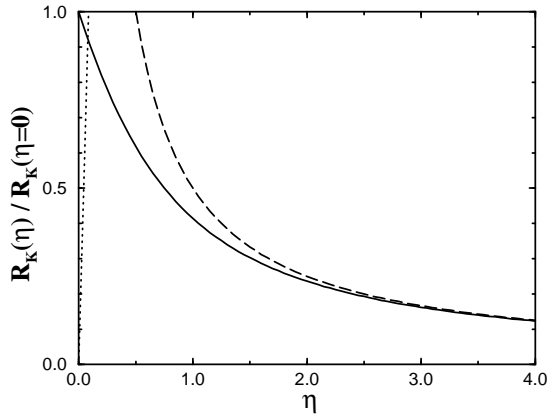, height=6cm}
 \caption{Kramers' correction factor to the rate as function of $\eta$:
 the solid and dotted curves correspond to the high (\protect\ref{ratetdep})
 and low (\protect\ref{kralowvis}) viscosity limit (the latter is
 calculated for a the barrier height of $E_b=5.8 ~\text{MeV}$,
 for $^{224}$Th at $1~\text{MeV}$) and
 the  dashed ones to the overdamped limit (\protect\ref{kraovda})}
\label{figure1}
\end{center}
\end{figure}

Notice please that the dependence of the rate on the effective
damping strength $\eta_b$ still is given by (\ref{ratetdep}).
Using the relation $\varpi^2=|C|/M$ (see (\ref{defgagaom})) the
limiting value at zero damping $R_K^{h.v.}(\eta_b=0)$ may be
written in the two equivalent forms
\begin{equation}\label{krahvzero} R_K^{h.v.}(\eta_b=0)=
\frac{\varpi_a}{2\pi} \,\sqrt{\frac{M_a}{M_b}}\exp(-E_b/T)
\equiv\frac{\varpi_b}{2\pi}\,\sqrt{\frac{C_a}{|C_b|}} \exp(-E_b/T)
.\end{equation} The influence of dissipation is visualized by
plotting in Fig.\ref{figure1} the ratio
$R_K^{h.v.}(\eta_b)/R_K^{h.v.} (\eta_b=0)$.
In addition to the result of formula (\ref{ratetdep}) we also show
two other cases. Firstly, we show a simplified version of the {\it
low viscosity limit}, \begin{equation}\label{kralowvis}
\frac{R_K^{l.v.}(\eta_b)}{ R_K^{h.v.}(\eta_b=0)}\;=\;2
\eta_b\frac{E_b}{T}\ . \end{equation} It is valid for very small
viscosity only (see eq.(\ref{condKdamp}) below) and provided
\cite{hantalbor} the action for the motion on top of the barrier
may be approximated by $I(E_b) \approx E_b/\varpi_b$. As
demonstrated in \cite{hitrans}, for the nuclear case such a
situation is found only at very small temperatures, much below the
critical temperature for pair correlations to become important.
Secondly, we explicitly indicate the limit the ratio
(\ref{ratetdep}) takes on for overdamped motion
\begin{equation}\label{kraovda} \frac{R_K^{h.v.}(\eta_b)}{
R_K^{h.v.}(\eta_b=0)}\;=\;\frac{1}{2 \eta_b} \qquad
\text{for}\qquad \eta_b \gg 1\;. \end{equation}

Whenever the "high viscosity limit" applies the influence of
dissipation manifests itself in a reduction of the decay rate over
the value given by (\ref{krahvzero}). This deviation is claimed to
allow for deducing a possible temperature dependence of
dissipation through the "measurable" rate. It must be noted,
however, that for {\it overdamped motion it is} not {\it the
effective damping factor $\eta_b$ which one deduces}. Indeed,
overdamped motion is governed by the {\it Smoluchowski equation in
which no inertia appears} (see Appendix \ref{implvarin}). But the
latter not only is present in $\eta_b$ but in $
R_K^{h.v.}(\eta_b=0)$ as well. A better way of writing the rate
formula in this case is
\begin{equation}\label{smolrate} R_{\rm ovd}=R_K^{h.v.}(\eta_b \gtrsim
1) = \frac{1}{2\pi}\, \sqrt{\frac{C_a}{|C_b|}}
\frac{|C_b|}{\gamma_b} \exp(-E_b/T) \equiv
\frac{1}{2\pi}\,\sqrt{\frac{C_a}{|C_b|}}
\frac{1}{\tau^b_{\text{coll}}} \exp(-E_b/T)\,. \end{equation}
Here, the time scale $\tau^b_{\text{coll}}$ appears which is
relevant for overdamped motion across the barrier, see below. As
can be inferred from Fig.\ref{figure1} this limit is actually
given for values of $\eta_b$ just above unity. Notice, please,
that it is only with the additional factor $\sqrt{M_a/M_b}$
included in (\ref{krahiviin}), on top of Kramers' classic version,
that the inertia drops out in the overdamped limit.

A few comments are in order on the validity of the rate formulas
of the high viscosity limit, for which the following assumptions
must hold true:
\begin{itemize}
  \item On the way from the minimum to the barrier the
  temperature must not change.
  \item The barrier must be sufficiently pronounced, first of all
  in the sense that its height be large as compared to the temperature, viz
  $ E_b\gg T $, for further details see sect.\ref{decravarin}.
  \item The effective damping rate must not be too small
 \begin{equation}\label{condKdamp} \eta_b \ge \frac{T}{2E_b}\,, \end{equation}
 otherwise formula (\ref{kralowvis}) would have to be applied.
\end{itemize}

It may be quite a delicate matter to fix or calculate the
temperature which is at stake here. For instance, a temperature
$T_{CN}$ associated to the {\it total} available energy for the
{\it compound nucleus} might be much larger, as for high initial
thermal excitations the system may cool down by emissions of
neutrons or $\gamma$'s before it fissions. Finally, we should like
to remark once more that presently any possible quantum features
are discarded, which might show up at low temperatures
\cite{hitrans}.

\section{Microscopic transport coefficients}\label{mictranspo}

Evidently, the temperature dependence of the rate will greatly be
influenced by that of the transport coefficients --- on top of the
influence through the Arrhenius factor $\exp(-E_b(T)/T)$. Let us
first look into results obtained applying linear response theory
within the locally harmonic approximation \cite{hofrep}, before we
turn to discuss other forms used in the literature, like in
\cite{dioszegi}. In this theoretical approach the transport
coefficients of average motion are obtained by relating in the low
frequency regime the strength distribution of a microscopically
calculated response function $\chi_{qq}(\omega)$ to the one of the
damped oscillator. The latter is defined as
\begin{equation}\label{defoscres} \left(\chi_{\text{osc}}
(\omega)\right)^{-1}\,q(\omega) \equiv -\left(M\omega^2 + i\gamma
\omega - C\right)\,q(\omega)= -q_{\rm ext}(\omega) \,.
\end{equation} and thus may be obtained by adding to (\ref{diffeq})
the term $-q_{\rm ext}(t)$ on the right and performing a Fourier
transformation. For overdamped motion the response function turns
into
\begin{equation}\label{resovdam} \chi_{\rm
ovd}(\omega)=\frac{i}{\gamma}\,\frac{1}{\omega +
iC/\gamma}\,.\end{equation} In accord with the remarks from above
on the Smoluchowski limit no inertia appears anymore.

This approach permits one to calculate the transport coefficients
as functions of shape and temperature for any given nucleus. The
formulation is done in such a way that on top of shell effects and
pairing (see \cite{ivhopair} with references to previous works)
collisional damping is accounted for as well (for a review see
\cite{hofrep}). As one may imagine, such computations are quite
involved, last not least because much knowledge is required about
various aspects of the dynamics of complex nuclear systems. This
is one of the reasons why as yet numerical computations have been
done only for particular nuclei or for more schematic cases
\cite{yahosa} - \cite{casta}, \cite{ivhopair}. Nevertheless, this
experience may allow us to deduce some gross features which may be
considered generic to a wider class of nuclear systems. This is
what we are going to do below. It seems appropriate, however, to
first add some general remarks concerning calculations based on
the deformed shell model as an approximation to the general mean
field.

The output of calculations of the type just mentioned contains
much more detailed information than at present one may possibly
relate to observable quantities. The coordinate dependence of the
transport coefficients, for instance, is one prime example. Often
in nuclear transport theories one simply has aimed  at constant
coefficients for inertia and friction. If calculated within the
linear response approach, on the other hand, sizable variations
with shape are seen. One may recall that a similar feature is
already seen in the potential landscape, when calculated with the
Strutinsky procedure, for instance. Besides the maxima and minima
which are typical for gross shell effects one sees detailed fine
structure. Such features may depend on peculiarities of the
underlying shell model, and may thus be unphysical in nature
already by that reason. For the dynamic transport coefficients
themselves further implications arise from quasi-crossings of
levels. To large extent such effects can be expected to become
much weaker in a multidimensional treatment, which at present
still is infeasible.

One should not forget that problems of this type are intimately
related to the fact that transport coefficients of inertia,
stiffness and friction are those of {\it average motion,
calculated on the level of the mean field}. Finally, however, they
are needed for an equation of motion of Fokker-Planck type which
accounts for dynamical {\it fluctuations}. The latter will help to
smooth out the variations of the transport coefficients in most
natural way. Evidently, the problem at stake here reflects the
general deficiency of mean field theory. In a more appropriate
treatment one would be able to treat self-consistently both the
mean field as well as its fluctuations. Since such a theory is not
available we suggest some other, more pragmatic procedure. As
described previously already, see e.g. \cite{ivhopair}, one may
smooth static energies as well as the other transport coefficients
with respect to their dependence on deformation. The averaging
interval in $Q$ is to be chosen large enough to wash out the rapid
fluctuations but small enough to preserve gross shell structures.

\subsection{Shell effects on potential landscape}
\label{shelpot}

\begin{figure}
\begin{center}
\epsfig{figure=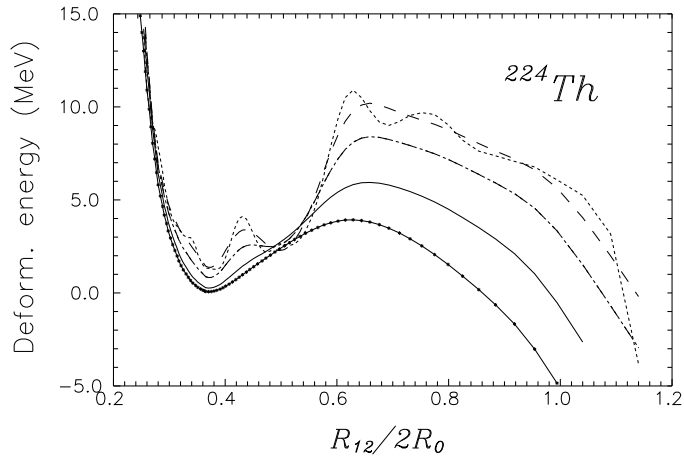, height=6cm}
 \caption{The deformation dependence of the collective potential energy.
The dotted curve shows the deformation energy at $T=0$. The
dashed, dot-dashed, solid and solid with stars curves correspond
to the averaged deformation energy at temperatures $T=0, 1, 2$ and
$3~\text{MeV}$. The deformation parameter here is the distance
$R_{12}$ between the centers of mass of left and right parts of
nucleus (devided by the diameter $2R_0$ of the sphere with equal
volume). The averaging is carried out on the interval $\Delta
(R_{12}/R_0) = 0.1$ . }
 \label{figure2}
\end{center}
\end{figure}

The static energy is calculated in the usual way as sum of a
liquid drop part and the shell correction, both of which depend on
temperature, for details see \cite{ivhopaya}. An example of the
deformation dependence of this potential energy is shown in
Fig.\ref{figure2}. The dotted curve represents the case of zero
thermal excitation. On top of the typical gross shell structure
fluctuations of smaller scale are recognized as well. Features of
this type lead to the rapid variations of the transport
coefficients we talked about above; for the local stiffness as the
second derivative of the static energy this is immediately
evident. As mentioned, we consider such fluctuations as
unphysical, for which reason we like to remove them by averaging
over an appropriate interval $\Delta Q$ but to keep the gross
shell structure. For the free energy, for instance, the smoothing
can be done in the following way: \begin{equation}\label{defavr}
\big\langle {\cal F}(Q,T) \big\rangle_{\text{av}}=\frac{\sum_i
{\cal F}(Q_i,T) f_{\text{av}}(\frac{Q-Q_i}{\Delta Q})} {\sum_i
f_{\text{av}} (\frac{Q-Q_i} {\Delta Q})}
\end{equation}
The smoothing function $f_{av}(x)$ in (\ref{defavr}) is taken to
be that of the Strutinsky shell correction method and the $Q_i$
are some points in deformation space. The use of the Strutinsky
smoothing function guarantees stability of the averaging
procedure: the smooth component of the deformation energy is
restored after smoothing again. This implies that the liquid drop
part of the energy is unchanged by this averaging.

In the figure we also show the averaged potential corresponding to
temperatures $T=0, 1, 2$ and $3 ~\text{MeV}$. As to be expected,
with increasing temperature the deformation energy becomes much
smoother and the height of the fission barrier gets reduced. This
is due to the reduction of shell effects, as well as to the
temperature dependence of the liquid drop part. At temperatures
above $T\approx 3 ~\text{MeV}$ the shell effects have disappeared
completely and the averaged deformation energy coincides with its
liquid drop component. As seen from the figure, at smaller
temperatures the shell correction, albeit averaged, does
contribute to the deformation energy and, hence, to the stiffness.
For example, at $T=1 ~\text{MeV}$ the stiffness at the barrier
(maximum of the deformation energy) is still several times larger
than that of the liquid drop part.

It should be mentioned that in \cite{yaivho} a somewhat different
(averaging) procedure was used. There the deformation energy was
approximated by two parabolas and the stiffness (at the minimum
and the barrier) was defined by the curvature of these parabola.
In this way shell effects are washed out to larger extent, not
only with respect to the fine structure but even with respect to
gross shell features. Consequently, the stiffness defined in this
way is rather close to the liquid drop stiffness.

\subsection{Temperature dependence of transport coefficients}
\label{tdepmicro}

\subsubsection{Local stiffness}\label{locsti}
In the following we identify stiffness as the one corresponding to
the free energy. Then we may write
  \begin{equation}\label{stiffness} C(T) = C_{\text{LDM}} + \delta C(T)\end{equation}
  Here, $\delta C(T)$ represents the contribution from the shell
  correction, which disappears with increasing temperatures. To
  parameterize the latter feature we take over a formula of
  \cite{bm} to get
  \begin{equation}\label{shdamptembm} \delta C(T) = \delta C(T=0) \frac{\tau}{\sinh
  \tau} \end{equation} with the shell correction parameter
  \begin{equation}\label{deftau} \tau = 2\pi^2 \frac{T}{\hbar
  \Omega_0} \qquad {\rm  and} \qquad
  \hbar \Omega_0 = \frac{41~\text{MeV}}{A^{1/3}}\end{equation} being the average shell spacing.
Above $\tau_{\rm shell}\simeq 3-5$, which corresponds to a
temperature of the order of \begin{equation}\label{shellT} T_{\rm
shell}
  \simeq (3-5)\frac{\hbar \Omega_0}{2\pi^2}\;\simeq\;1-2 ~\text{MeV}\,,
  \end{equation} the $\delta C(T)$ practically vanishes, such
that $C(T)$ attains its liquid drop value. Eventually, this
$C_{\text{LDM}}$ may still be treated as $T$-dependent
\cite{gustrubra}. Finally, we should recall our suggestion from
above to average the transport coefficients over smaller intervals
in $Q$. In this sense the $\delta C(T=0)$ is meant to only
represent gross shell features.

\subsubsection{Local inertia} \label{inert}

As mentioned already in the Introduction, one should expect the
inertia to vary with temperature. It is more than tempting to
assume a form similar to the one for the stiffness, namely
\begin{equation}\label{inertia} M(T) = M_{\text{LDM}} + \delta
M(T)\end{equation} in which the last term drops to zero like given
in (\ref{shdamptembm}). Indeed, within the linear response
approach a behavior of that type has been observed in a numerical
study \cite{hoyaje}. There, the value reached at larger
temperatures was given by that of irrotational flow, which for the
present notation means to identify $M_{\text{LDM}}=M_{\rm irrot}$.
To the best of our knowledge, there is no other theoretical model
where such a transition is seen explicitly --- although one must
say that in phenomenological applications of transport models
commonly the $M_{\rm irrot}$ is taken to represent the macroscopic
value of inertia. At present the conjecture behind (\ref{inertia})
still lacks a direct and general proof. However, in
\cite{magvydhof} the nucleonic response function has been studied
applying Periodic Orbit Theory (POT). There it was seen that its
"fluctuating part" $\delta \chi(\omega)$ decreases with $T$ like
the shell correction to the static energy. For slow collective
motion the inertia is determined by the second derivative of this
response function with respect to frequency calculated at
$\omega=0$ (see App.\ref{schemod} and eq.(\ref{inert1m}) in
particular). Therefore, within such a model the "shell correction"
to the inertia was indeed proven to behave as claimed above,
although several questions remain open. Amongst others, it is
unclear to which extent this proof would get modified after
considering "collisional damping". The latter cannot be treated
within POT but should play a major role for the transition to
hydrodynamic behavior. Possible reasons for rendering a
microscopic approach quite difficult have been reported in
\cite{yaivho, ivhopaya, hofrep}. On the microscopic level they are
related to the strength distribution for the local collective
motion. The liquid drop model, on the other hand, represents
motion of a system having a sharp surface, in contrast to
microscopic calculations involving the diffuse surface of the mean
field and, hence, of the density, for details see \cite{ivhopaya}.
Fortunately, however, at larger $T$ when the collisions become
more and more important the motion gets strongly damped such that
the inertia drops out anyway, see (\ref{resovdam}).

Because of these difficulties in microscopic computations we
propose to fix the $M(T)$ through the vibrational frequency
$\varpi$ and the local stiffness by the relation given in
(\ref{defgagaom}), namely $M=|C|/\varpi^2$. For our version of
Kramers' rate formula this was easily achieved in using the second
variant shown in (\ref{krahvzero}).

\subsubsection{Vibrational frequency} \label{vibfreq}

\begin{figure}
\begin{center}
\epsfig{figure=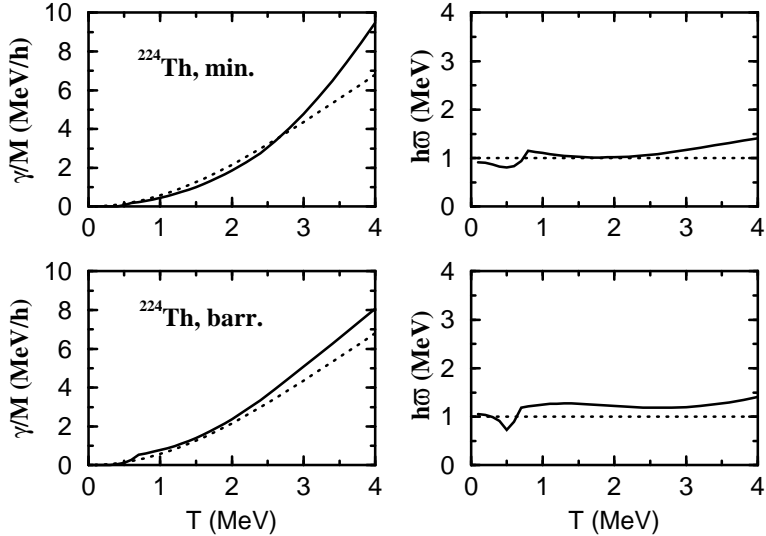, height=8cm}
 \caption{The inverse relaxation time
 $1/\tau_{\text{kin}}=\Gamma_{\text{kin}}/\hbar$
 (left panel)  and  $\hbar\varpi$ (right panel) as function of
 temperature: the microscopic results (solid curves) compared to the
 approximations  (\protect\ref{choifreq}) and
 (\protect\ref{gamkinap}) (dotted curves). }
 \label{figure3}
\end{center}
\end{figure}

At the extremal points of the potential landscape this frequency
$\varpi$ is a well defined quantity. To some extent it is even
accessible to experimental verification, at least for zero thermal
excitation. At the minimum it may be associated with the energy of
a collective mode (for very recent work on this subject see
\cite{thirolf}) and for the barrier it influences the
penetrability, as encountered for instance in neutron induced
fission \cite{bjolyn}. Generally, the $\hbar\varpi$ is believed to
be of the order of $1 ~\text{MeV}$. Indeed, numerical calculations
for $^{224}$Th \cite{yaivho, ivhopaya} show this to be quite
insensitive to temperature; to lesser extent this is true also for
the variation with shape and mass number. Altogether, for a first
orientation the following choice seems appropriate
\begin{equation}\label{choifreq} \hbar\varpi_a \simeq
\hbar\varpi_b \simeq 1 ~\text{MeV}\,
\end{equation} with deviations being within a factor of 2 or less. This
appears to be the case even when pairing is included at smaller
$T$. In Fig.\ref{figure3}
we take up the case of $^{224}$Th, again. The calculation is the
same as reported in \cite{ivhopair}; more details will be given
below in sect.\ref{infpair}. From the right panel it is seen that
this conjecture is pretty much fulfilled.

\subsubsection{Ratio of friction to inertia} \label{ratfriin}

As said above, see eq.(\ref{defgagaom}), the ratio $\gamma/M$
determines the inverse relaxation time to the Maxwell
distribution. For underdamped motion this quantity also defines
the width $\Gamma_{\text{kin}}$ of the strength distribution.  In
Fig.\ref{figure3} we show it on the left hand panel as function of
$T$. The dashed curve represents the following approximation,
details of which are discussed in App.\ref{schemod}, namely
\begin{equation}\label{gamkinap}
\frac{\gamma}{M}\hbar=\Gamma_{\text{kin}} \approx 2\Gamma_{\rm
sp}(\mu,T) = \frac{2}{\Gamma_0} \;\frac{\pi^2T^2}{1+ \pi^2T^2/c^2}
\approx \frac{0.6T^2}{1+ T^2/40} ~\text{MeV} \qquad (T ~{\rm in}
~\text{MeV})\,.
\end{equation} As expected it represents the microscopic result
quite well at smaller values of $T$ which correspond to smaller
values of damping. Recall, please, that the overdamped limit is
given already for values of the damping factor $\eta_b$ slightly
above $1$, see Fig.\ref{figure1}.. As can be inferred from
Fig.\ref{figure3} and the estimate (\ref{choifreq}), this happens
at temperatures above $2~\text{MeV}$; mind that
$\eta=(\gamma/M)(2\omega)^{-1}$.

\subsubsection{Ratio of friction to stiffness} \label{ratfristi}

\begin{figure}
\begin{center}
\epsfig{figure=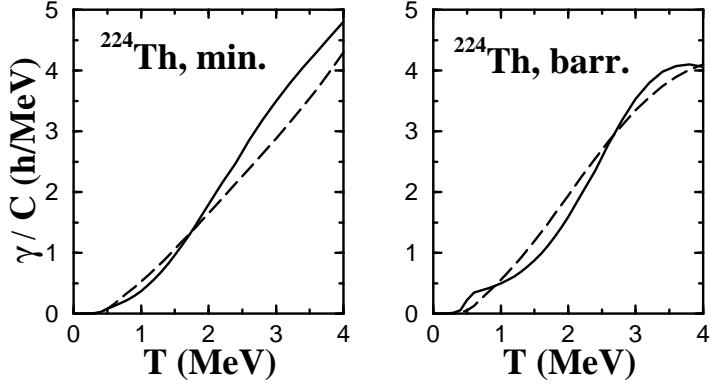, height=6cm}
 \caption{The relaxation time $\tau_{\text{coll}}$ for collective motion at
 the potential minimum and at the barrier: the microscopic result (solid
curve) compared to the approximation
(\protect\ref{relcoll}-\protect\ref{cmacro}) (dotted curve).}
 \label{figure4}
\end{center}
\end{figure}

In Fig.\ref{figure4} we plot, as function of $T$, the time
$\tau_{\text{coll}} = \gamma/|C|$, which measures the local
relaxation in the coordinate.  We may recall from
(\ref{defgagaom}) that for the overdamped case this is the only
relevant time scale left. Its inverse determines the width of the
strength distribution along the imaginary axis (see
(\ref{resovdam})). Likewise the decay rate (\ref{smolrate})
associated to the Smoluchowski equation is proportional to
$\tau_{\text{coll}}^{-1}$.
In Fig.\ref{figure4}, again, the fully drawn line shows the
microscopic result. The dashed curves represent an approximation,
into which the following two features are incorporated, the
decrease of the stiffness (as given in (\ref{stiffness}) and
(\ref{shdamptembm})) and the fact that with increasing $T$ the
friction coefficient reaches a plateau \cite{hoivya, yaivho}. To
combine both effects we chose a functional form similar to the one
for the $\Gamma_{\text{kin}}$ of (\ref{gamkinap}) (see
App.\ref{schemod}) but with a different cut-off parameter $c_{\rm
macro}$:
\begin{equation}\label{relcoll}
\tau_{\rm coll}=\frac{\gamma}{|C|}\approx
\frac{2}{\hbar\varpi^2\Gamma_0} \;\frac{\pi^2T^2}{1+
\pi^2T^2/c_{\rm macro}^2}\approx\frac{0.6T^2}{1+ \pi^2T^2/c_{\rm
macro}^2}\frac{\hbar}{{\text{MeV}}} \qquad (T, c_{\rm macro} ~{\rm
in} ~\text{MeV})\,, \end{equation} with $\hbar\varpi\approx 1 $
MeV. One should expect the $\tau_{\text{coll}}$ to reach a
macroscopic limit like \begin{equation}\label{cutomac}
\left.\tau_{\text{coll}}\right|_{T_{\rm
h.T}}=\left.\frac{\gamma(T)}{|C(T)|}\right|_{T_{\rm h.T}}\approx
\frac{\gamma_{\rm wall}/2}{|C_{\text{LDM}}(T)|}\,,\end{equation}
at larger temperatures. With a parameterization as in
(\ref{relcoll}) the limit is obtained above $T_{\rm h.T}\simeq
c_{\rm macro}/\pi$, for which reason the $c_{\rm macro}$ would be
given by
\begin{equation}\label{cmacro}c^2_{\rm macro}=
\frac{\hbar\varpi^2\Gamma_0}{2}\frac{\gamma_{\rm wall}/2}{|C_{\rm
LDM}(T)|}\approx 8.2 \frac{\gamma_{\rm wall}}{|C_{\rm
LDM}(T)|}\frac{\text{MeV}^3}{\hbar}\,.\end{equation} Here, we
accounted for results obtained by several previous numerical
calculations, see e.g.\cite{yaivho, ivhopaya, hofrep}. They showed
that the value of friction at large $T$ is somewhat below the wall
formula. The factor $1/2$ is only to be considered a rough rule of
thumb. For the stiffness, on the other hand, the macroscopic limit
evidently is given by the liquid drop model. As the microscopic
calculation was done with a $T$-dependent $|C_{\text{LDM}}(T)|$ we
chose the same one in this fit. In both curves the effects of
pairing were included, which we are going to address now.

\subsubsection{The influence of pairing} \label{infpair}

\begin{figure}
\begin{center}
\epsfig{figure=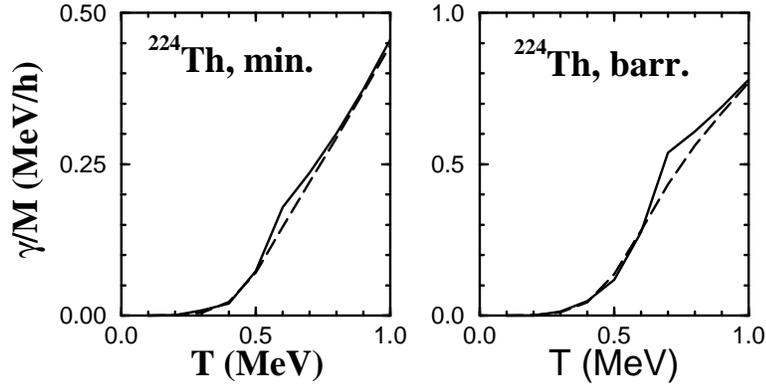, height=6cm}
 \caption{The influence of pairing on the inverse relaxation time
 $\tau_{\text{kin}}^{-1}=\Gamma_{\text{kin}}/\hbar$.
The microscopic results of \protect\cite{ivhopair} are compared
with approximation (\protect\ref{modpairgm}).}
 \label{figure5}
\end{center}
\end{figure}

This problem has recently been studied in \cite{ivhopair}. The
fully drawn lines shown in the previous figures refer to such a
calculation. Whereas in \cite{ivhopair} one concentrated on the
regime in which pairing is expected to be effective, the present
results extend up to $T=4~\text{MeV}$. Calculations in that regime
had been reported before in \cite{ivhopaya}. The underlying shell
model is the same in both cases, but a different procedure is
applied for the single particle width $\Gamma_{\rm sp}$. For the
unpaired case the form (\ref{spwidth}) was used, for which the
frequency dependence of $\Gamma_{\rm sp}(\omega,T)$ leads to
convolution integrals in the response functions. They are known to
reduce the collective widths \cite{hofrep}. In the paired case
such a calculation is no longer feasible, for which reason there a
constant $\Gamma_{\rm sp}(\mu,\Delta,T)$ had been assumed, with
$\Delta$ being the pairing gap. To have a more or less smooth
transition to the unpaired case, we now approximate the
$\Gamma_{\rm sp}(\omega,T)$ by the $\Gamma_{\rm sp}(\mu,\Delta,
T)$  which above the critical temperature for pairing reduces to
the $\Gamma_{\rm sp}(\mu,T)$ given in (\ref{widthspwi}). For this
reason our present friction coefficient may be overestimated
slightly. For $\Gamma_{\text{kin}}/\hbar=\gamma/M$ the new results
are shown in Fig.\ref{figure5},
where we concentrate on temperatures up to $1~\text{MeV}$. To
simulate the apparent effect of pairing to reduce friction we
suggest the modified formulas \begin{equation}\label{modpairgm}
\frac{\gamma}{M} \,= \,f_{\rm pair}\,
\frac{\gamma(\Delta=0)}{M(\Delta=0)}\,\equiv \,f_{\rm pair}\,
\frac{\Gamma_{\text{kin}}(\Delta=0)}{\hbar}
\end{equation}
and \begin{equation}\label{modpairgc} \frac{\gamma}{|C|}
=\tau_{\text{kin}}\, = \,f_{\rm pair}\,
\tau_{\text{kin}}(\Delta=0)\,. \end{equation} Here, $f_{\rm pair}$
parameterizes the decrease of friction due to pair correlations:
An ansatz like  \begin{equation}\label{paircor} f_{\rm pair}=
\frac{1}{1+ \exp( -a(T-T_0))}\;,\end{equation} may do with the
following parameters: $ a =10 ~\text{MeV}^{-1}$ and $ T_0 =0.55
~\text{MeV}$ at the barrier, and $ a =12 ~\text{MeV}^{-1}$, $T_0 =
0.48 ~\text{MeV}$ at the minimum. It was found that this choice
fits best the microscopic results (with the functional form
(\ref{modpairgm})). It is seen that these values vary with shape,
which may be an indication that they are perhaps different for
different nuclei. For a first orientation such details might be
discarded. Then an average of the two values could be used both
for $a$ and $T_0$. Finally, we should like to remark that this
influence of pairing is most dramatic for friction, and much less
so for $M$ and $C$. Therefore, we suggest to neglect the influence
on the latter.

 \section{Temperature dependent decay rates} \label{temprate}

If one wants to gain information on the transport coefficients and
their $T$-dependence, in particular, one needs to separate the
influence of the pre-factors from the more or less trivial
exponential part (the "Arrhenius factor"). Commonly, this feature
is then simply identified by Kramers' conventional factor which
only depends on $\eta_b$. A systematic study performed in
\cite{thoebert} has revealed the appearance of a {\em threshold
temperature} $T_{\text{thresh}}$ above which deviations from the
statistical model are seen over a wide range of fissioning
systems. Moreover, its ratio over the temperature dependent
barrier heights $E_{\text{Bar}}(T)\equiv E_b(T)$ showed a
remarkable insensitivity on mass number A. For later purpose it is
more convenient to divide that ratio by two to get
\begin{equation}\label{threshTB} \left. \frac{T}{2E_b(T)}\right|_{\text{thresh}}
\simeq 0.13 \qquad \text{from Fig. 4 of \cite{thoebert}}.
\end{equation} We are now going to view this result in the light
of our microscopic transport coefficients. At first we shall
concentrate on the ratio $R_K^{h.v.}(\eta_b)/R_K^{h.v.}
(\eta_b=0)$, to comment later on the $T$-dependence of the
additional factors seen in (\ref{krahvzero}), and which involve
ratios of the inertias or stiffnesses.

\subsection{Rate from microscopic transport
coefficients}\label{ratemicro}

\begin{figure}
\begin{center}
\epsfig{figure=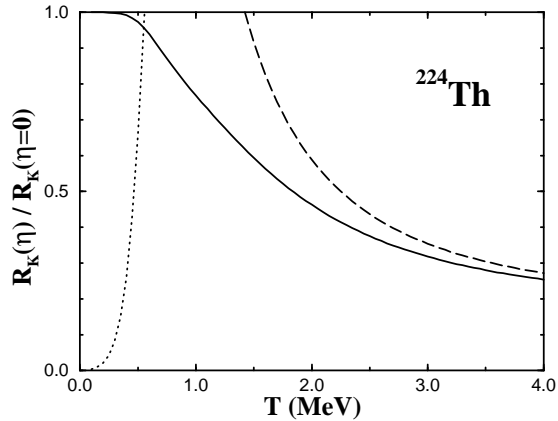, height=6cm}
 \caption{Temperature dependence through a microscopic
 $\eta=\eta(T)$ (fully drawn line); the dashed curve represents
 the overdamped case as given by (\protect\ref{kraovda}), the dotted one
 corresponds to the low viscosity limit as given by
 (\protect\ref{kralowvis}) for the same barrier height as in
 Fig.(\protect\ref{figure1}).}
  \label{figure6}
\end{center}
\end{figure}

In Fig.\ref{figure6} we plot the normalized rates for the three
cases shown already in Fig.\ref{figure1}, but now as function of
temperature as determined by our microscopic $\eta(T)$.
Various conclusions may be drawn from this figure:
\begin{itemize}
  \item Above $T\simeq 0.6-0.7 ~\text{MeV}$ an "onset of dissipation" is seen,
   indeed, in the sense of a decrease of the rate for the "high viscosity"
   limit. The value of this "threshold temperature" is in reasonable
  agreement with the $T_{\text{thresh}}\simeq 1.09 - 1.22$ given
  in Tab.1 of \cite{thoebert} for $^{224}$Th.
  \item Moreover, from the right panel of Fig.\ref{figure5}
  together with the value for the frequency given in Fig.\ref{figure3}
  one may deduce for these temperatures the $\eta_b$ to be larger
  than the value of $T/2E_b$ given in (\ref{threshTB}). It then
  follows from (\ref{condKdamp}) that the deviation from the
  statistical model is to be attributed to the high viscosity limit.
  \item This feature evidently is related to the disappearance
   of pair correlations at $T\simeq 0.5 ~\text{MeV}$.
   The curve for the "low viscosity limit" demonstrates
  that below this temperature the former case should not be
  used at all (mind condition (\ref{condKdamp})).
  \item As indicated earlier, above $T\simeq 2~\text{MeV}$  the overdamped limit
  applies, for which we suggested to use formula (\ref{smolrate}).
\end{itemize}

\subsection{Comparison with phenomenological models}\label{comphemo}

\begin{figure}
\begin{center}
\epsfig{figure=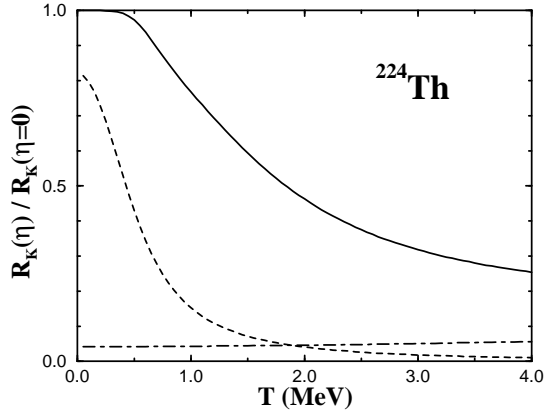, height=6cm}
 \caption{Temperature dependence of the decay rate in high viscosity
 limit: Fully drawn curve: same as in Fig.\protect\ref{figure6};
 dashed curve: $\eta_b$ given by (\protect\ref{phenTdepetaI}); dashed-dotted
 curve: wall friction and liquid drop values for stiffness
 and inertia.}
 \label{figure7}
\end{center}
\end{figure}

In Fig.\ref{figure7} we compare the result shown in the previous
figure (for formula (\ref{ratetdep})) with the one of a
macroscopic picture, together with that suggested in
\cite{backetal, dioszegi}. By the macroscopic model we mean to use
the wall formula for friction, the liquid drop model for inertia
(irrotational flow) as well as for the stiffness. In
\cite{backetal} the following functional form for $\eta_b(T)$ had
been suggested:
\begin{equation}\label{phenTdepetaI} \eta(T) = 0.2 + 3
T^2\end{equation}
It is seen that for such a $T$-dependence the  onset of
dissipation occurs quite abruptly and at rather small
temperatures. One must have in mind, though, that the ratio
plotted in Fig.\ref{figure7} is sensitive to the temperature at
the barrier, as it is this one which determines the $\eta_b$. This
fact might explain why the transition exhibited here through the
dashed curve lies at smaller $T$ than discussed in \cite{backetal,
dioszegi}. It can be said that for the other ansatz $ \eta(T) =
0.2 + 5 T$ used as well the transition would occur at even smaller
temperatures. The reason seems obvious: In these two forms {\it
the dissipation rate $\eta$ is much too large at small $T$}
--- at least too large as compared to our microscopic results.
Indeed, as discussed above, at small temperatures {\it pairing
correlations require dissipation to vanish}.

\subsection{Relation to the statistical model}
\label{relStaMo}

The rate of the high viscosity limit when extrapolated down to
zero friction, $R_K^{h.v.}(\eta_b=0)$, corresponds to a variant of
the transition state method, and thus to one \cite{kram} of the
Bohr-Wheeler formula \cite{bohrwheel} (see also \cite{strutkram}).
We demonstrate in sect.\ref{rate-trans} that this still holds true
for the modified version of variable inertia. One must have in
mind, however, that such an extrapolation may be meaningless as,
by its very construction, the transition state method bases on the
assumption of a complete equilibrium. The latter is given only if
sizable friction forces provide sufficiently fast relaxation to
this global equilibrium. Our result for the temperature dependence
of friction then imply that any version of the transition state
result must be taken with reservation when applied at small
thermal excitations. At this point it would be too much to compare
complicated evaluations of the Bohr-Wheeler formula with estimates
of Kramers' rate which have our microscopic transport coefficients
as input. This must be subject of further studies.

What is feasible, however, is to compare our results with those of
a statistical model, in which the Bohr-Wheeler formula
\cite{bohrwheel} is approximated by
\begin{equation}\label{rate-stat} R_{stat} = \frac{T}{2\pi\hbar} \exp(-E_b/T)
\,, \end{equation} see e.g.\cite{sije, vandhuiz}. This approximate
form comes up when in the Bohr-Wheeler formula \cite{bohrwheel}
the level density at the barrier is identified as that of the
total system, whereas in the correct expression the collective
degree of freedom has to be excluded (see e.g.\cite{strutkram}).
That such an approximation may lead to erroneous results when
interpreting data has been pointed out also in \cite{thoenn}. We
may briefly follow up this discussion by using our microscopic
input.

\begin{figure}
\begin{center}
\epsfig{figure=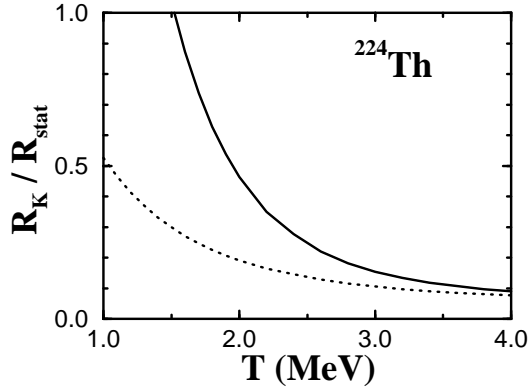, height=6cm}
 \caption{The ratio (\protect\ref{ra-staKRov}) of Kramers' rate
 to that of the statistical model (fully drawn line).
 The dotted curve demonstrates the influence of the stiffnesses,
 which now are calculated from the liquid drop model, but with friction
 unchanged. }
  \label{figure8}
\end{center}
\end{figure}

The ratio $R_K/R_{stat}$ may be calculated for the three cases we
looked at before, low and high viscosity limit as well as the
overdamped limit. For "high viscosity" one gets from
(\ref{krahiviin}) (mind (\ref{krahvzero})):
\begin{equation}\label{rate-statKR}
\frac{R_K^{h.v.}}{R_{stat}}=\frac{\hbar\varpi_b}{T}\,\sqrt{\frac{C_a}{|C_b|}}
 \left(\sqrt{1+\eta_b^2} - \eta_b\right)\,. \end{equation}
For overdamped motion, expression (\ref{smolrate}) leads to
\begin{equation}\label{ra-staKRov} \frac{R_K}{R_{stat}} =
\sqrt{\frac{C_a}{|C_b|}}
\frac{|C_b|}{\gamma_b}\frac{\hbar}{T}\,.\end{equation} The result
(\ref{rate-stat}) can be expected to deviate sizably from Kramers'
one. This is demonstrated in Fig.\ref{figure8} by the fully drawn
line which represents the ratio given in (\ref{rate-statKR}).
Here, the "onset of dissipation" seemingly is even more
pronounced, and the deviation starts at a much higher temperature.
However, {\it these effects are not only related to dissipation}.
Besides the more or less obvious fact of the prefactor in
(\ref{rate-stat}) being not identical to the frequency $\hbar
\varpi_a \simeq \hbar \varpi_b$, there appears the square root of
the ratios of the two stiffnesses. The latter is largely
influenced by shell effects, which are known to be sensitive to
variations of temperature. The implication of this feature on the
ratio $R_K/R_{stat}$ is exhibited in Fig.\ref{figure8} by the
dotted curve. Its deviation from the solid one is solely due to
the stiffnesses being evaluated from the ($T$-independent) liquid
drop model $C_{\text{LDM}}(T=0)$.

Obviously, the $R_{stat}$ becomes small at small temperatures, say
below about $0.5 ~\text{MeV}$. In this range nuclear friction is
small, too, not only at the barrier but also inside the well (see
Figs.\ref{figure3} and \ref{figure4}). Discarding any quantum
effects, which in this regime may become important \cite{hitrans},
one might then use Kramers' low viscosity limit. As mentioned
previously, any transition state result, however, does {\it not
apply, for which reason a comparison of both is meaningless}. For
the Bohr-Wheeler formula to be valid the system inside the well
has to be in {\it complete equilibrium} \cite{hantalbor}. Such a
situation is given only for sufficiently large damping.

\subsection{Isotopic effects} \label{isoshit}

\begin{figure}
\begin{center}
\epsfig{figure=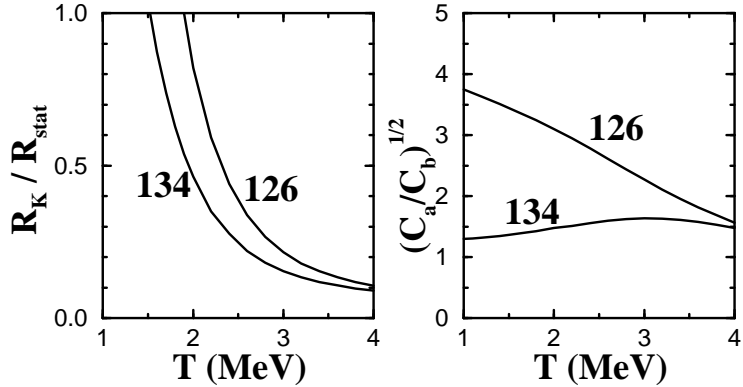, height=6cm}
 \caption{Results for different
 Th isotopes; left panel: Ratio of Kramers' rate to that of
 the statistical model, right panel: influence of shell
 structure through ratio of stiffnesses.}
  \label{figure9}
\end{center}
\end{figure}

In \cite{backetal} different nuclei have been studied
experimentally and analyzed with respect to a temperature
dependence of the dissipation strength $\eta_b$ (called $\gamma$
there). In particular two isotopes of Thorium have been examined,
namely $^{224}$Th and $^{216}$Th corresponding to neutron numbers
of $N=134$ and $126$, respectively. The different behavior seen in
experiment has been fully attributed to this $\eta_b$ and in this
way very different values of $\eta_b$ have been found, together
with a different increase with $T$. Evidently such features cannot
be explained within a macroscopic picture. One needs to account
properly for shell effects. From our experience with microscopic
computations it seems unlikely that they would have such a big
influence on friction alone. On the other hand, recalling that
$N=126$ corresponds to a closed shell, it is evident that the
stiffnesses for $N=134$ and $126$  will be quite different, in
particular for the ground state minimum. This effect can be seen
on the right panel of Fig.\ref{figure9}, which shows the square
root of the relevant ratios to be quite different for the two
isotopes.
Moreover, it is easy to convince oneself that this effect is in
the right direction. Indeed, as can be seen from
(\ref{rate-statKR}), smaller values of this ratio
$\sqrt{C_a/|C_b|}$ {\it simulate} larger values of $\eta_b$.
Notice, please, that an excitation of about $100~\text{MeV}$, as
given in Fig.9 of \cite{backetal}, corresponds to a temperature of
$T \approx 2~\text{MeV}$. Finally, on the left panel in
Fig.\ref{figure9} we show the full ratio as given by
(\ref{ra-staKRov}). Evidently, we are not able to fully explain
the results shown in Fig.9 of \cite{backetal}, but this would have
been asking too much, for various more or less obvious reasons.

\section{Summary and discussion} \label{sumdisc}

One of our main goals was to suggest simple forms of the
temperature dependence of those quantities which parameterize
transport in collective phase space. These suggestions are based
on simplified pictures for the intrinsic response combined with
experience in microscopic computations within realistic approaches
employing the deformed shell model. Rather than using the
transport coefficients themselves we argued in favor of
combinations which allow for a more direct physical
interpretation.

These results were tested at the fission decay rate. To make our
reasoning as transparent as possible, a simple one dimensional
model was used, for which Kramers' picture can be applied. This
model is somewhat schematic, in so far as the potential is assumed
to have one pronounced minimum and one pronounced barrier only,
which follow more or less closely the form of two oscillators, one
upright and one inverted. Different to the original work of
Kramers and of many subsequent applications, our transport
coefficients were allowed to vary along the fission path. In a
first approximation this leads to a modification of Kramers'
original rate formula. For variable inertia, such a modification
already appears necessary for the simple reason that for large
damping one should reach the rate formula corresponding to the
Smoluchowski equation. Still, the deviation from the undamped case
is solely determined by the damping strength $\eta_b$ calculated
at the barrier. However, for a complete understanding of
temperature effects of the decay rate it does not suffice to
concentrate only on this damping strength.

Of course, such a simple model for the static energy may not apply
at all at (smaller) temperatures when shell effects lead to
important deviations. It is unclear how one could generalize the
formulas of Kramers and Bohr-Wheeler to realistic cases with two
or perhaps three barriers having minima in between. To calculate
the decay rate for such potentials it is perhaps simpler to use
the more general description in terms of Fokker-Planck or Langevin
equations. It may then also be possible to account better for the
variations of the other transport coefficients with coordinate and
temperature, namely inertia and friction. To the best of our
knowledge so far in such calculations only the transport
coefficients of macroscopic models have been used. However, the
content of Fig.\ref{figure7} indicates that greater deviations can
be expected for microscopic inputs.

Nevertheless, within our model we are able to clarify a few
important points which must not be discarded even at temperatures
at which shell effects do not really dominate the process.

\begin{itemize}

\item The transition to overdamped motion already occurs at
temperatures around $T \simeq 2~\text{MeV}$. Then the decay rate
should be calculated from the Smoluchowski equation.

\item In this case no inertia appears any more. Thus neither the
frequency $\varpi$ nor the effective width $\Gamma_{\rm
kin}=\hbar\gamma/M$ (sometimes referred to as $\hbar \beta$) play
any role.

\item The solely important quantities are the ratio of friction
to stiffness, $\gamma/C$, and (for the decay rate) the ratio
$C_a/C_b$ of the stiffnesses. Physically, the former determines
the relaxation time for sliding motion in the potential, see
(\ref{relcoll}). In passing we may note that it is essentially
this time which determines the saddle to scission time, an effect
not considered here. Notice that these stiffnesses are subject to
large shell effects which may easily be accounted for by the shell
correction method. As a first demonstration of this effect we have
calculated the impact of these stiffnesses on the decay rate of
the Thorium isotopes $N=134$ and $N=126$.

\item In principle, the vibrational frequency $\varpi$ should be
obtained from microscopic computations (for $T<2 $ MeV).  But a
simple and quite fair estimate may be given by $\hbar\varpi_a
\simeq \hbar\varpi_b \simeq 1 ~\text{MeV}$ (\ref{choifreq}), which
is almost exact for the case of $^{224}\text{Th}$. Evidently, the
$\varpi$ may be greatly influenced by shell effects, but from our
experience we claim that the deviations are within a factor of
about two.

\item Finally, we should like to briefly comment on the paper
\cite{weidzhang}. There scaling rules have been derived for the
case of the Smoluchowski equation, and based on phenomenological
input. In the schematic model underlying the discussion a few
assumptions had been made which are not in accord with microscopic
results. First of all, the ratio of barrier height to temperature
definitely decreases with $T$. Secondly, as the authors monitor
themselves, a frequency at the barrier of the order of
$20~\text{MeV}$ is much too high. Together with the value used for
$\beta$, namely $\beta\equiv \Gamma_{\text{kin}}/\hbar \simeq
10^{22}~\text{sec}^{-1}$ one gets an $\eta_b$ of $ \simeq 0.2$.
This implies the motion around the barrier to be  {\it under}
damped rather than overdamped. Moreover, for several cases studied
in this paper, the $\eta_b$ is seen to be of the order of or even
smaller than $T/2E_b$. Thus, according to (\ref{condKdamp}) not
even Kramers' high viscosity limit seems appropriate.

\end{itemize}

Let us turn now to very low temperatures, say to the regime where
pairing correlations become important. They imply an additional
reduction of dissipation, such that one may truly speak of the
onset of dissipation when pair correlations disappear. It must be
said though that this transition occurs at temperatures definitely
smaller than those suggested in \cite{pauthoe} - \cite{strasb}. As
mentioned earlier, one has to make sure, however, that one speaks
of the same temperature. The one of the compound nucleus might be
larger than the one the system still has when it passes the
barrier.

It ought to be stressed that a small damping strength at small
temperatures may have quite drastic implications. In case the
"high viscosity limit" still applies quantum corrections to the
decay rate would lead to an increase of the latter \cite{hitrans}.
If, however, the dissipation strength falls below the limit given
by (\ref{condKdamp}) the nature of the diffusion process would
change completely. Then dissipation is too weak to warrant
relaxation to a quasi-equilibrium. This not only violates Kramers'
rate formula (for the high viscosity limit), or our extension of
it, but also the Bohr-Wheeler formula becomes inapplicable.
Moreover, so far no method exists of how one might incorporate
collective quantum effects.

Finally we should like to indicate that our transport coefficients
are not free of uncertainties in some of the parameters specifying
our microscopic input. The most difficult one is found in the
single particle width. It has been parameterized in \cite{jehosi}
by referring to the optical model for nucleons in finite nuclei
through a kind of local density approximation. There the
$\Gamma_0$ and its dependence on the nuclear density had been
traced back to microscopic computations of the self energies in a
generalized Brueckner type description. This implies an
uncertainty of perhaps a factor of two in all formulas where the
friction coefficient appears, as in (\ref{gamkinap}) or
(\ref{relcoll}) as well as in the $\eta$ of (\ref{etaforms}).
Since any microscopic answer on such questions is extremely
difficult, there is hope of narrowing down this uncertainty by
more elaborate comparisons with experimental results.

\appendix
    \makeatletter
    \@addtoreset{equation}{section}
    \def\theequation{\thesection.\arabic{equation}}
    \makeatother
\section{Implications from a variable inertia} \label{implvarin}

In Kramers' seminal paper \cite{kram} the equation for the density
in phase space was written and applied to the decay of a
metastable system for the case of constant transport coefficients.
In nuclear physics both the inertia as well as the friction
coefficient vary with the collective coordinate. This has been
accounted for already in early applications to heavy ion
collisions, where globally Gaussian solutions (centered at the
classical trajectories) were used to calculate reaction cross
sections see e.g. \cite{HOFNGO}. Written in compact form for one
variable the transport equation should read:
\begin{equation}\label{KramersQP}
\begin{split}\frac{\partial}{\partial t} \,f(Q,P,t) = &
\left[ -\frac{P}{M(Q)} \,\frac{\partial}{\partial Q} +
\frac{\partial}{\partial Q} \left( \frac{P^{2}}{2 M(Q)} + V(Q)
\right) \,\frac{\partial}{\partial P} \right.  \\ &   + \left.
\frac{\partial}{\partial P} \left(\frac{P}{M(Q)} \,\gamma(Q) +
D(Q) \,\frac{\partial}{\partial P} \right)\right] f(Q,P,t)\,.\\
\end{split}\end{equation} Neglecting any quantum effects, which
might show up at small temperatures only \cite{hofrep, hitrans},
the diffusion coefficient is given by the classic Einstein
relation $ D(Q)=\gamma(Q) T$. A possible $T$-dependence of the
transport coefficients has not been indicated explicitly. This is
immaterial as long as we treat temperature as a fixed parameter,
which is assumed to hold true in the entire paper. As is easily
verified, one stationary solution of (\ref{KramersQP}) is the
distribution of global
equilibrium,\begin{equation}\label{equidensQP} f_{\text{eq}}(Q,P)
= \frac{1}{Z} \,\exp (-\beta \mathcal{H}(Q,P))\,
\end{equation} with the energy given by the classical Hamilton function
 \begin{equation}\label{HamiltonQP} \mathcal{H}(Q,P) = \frac{P^{2}}{2 M(Q)} + V(Q)\,.
\end{equation} The reason is due to the following
features: (i) The conservative part of the equation has the form
of Liouville's equation, namely \begin{equation}\label{Liouville}
\frac{\partial}{\partial t} \,f(Q,P,t) = \Bigl\{\mathcal{H}(Q,P)
\,, \, f(Q,P,t)\Bigr\} \,.\end{equation} (ii) The terms which
represent dissipative and fluctuating forces are assumed
independent of the momentum $P$, such that $P$ only appears
quadratically in the kinetic energy. (iii) The choice of the
diffusion coefficient by $ D(Q)=\gamma(Q) T$ makes the second line
of (\ref{KramersQP}) vanish once applied to (\ref{equidensQP}).

The structure of these equations not only allow for the proper
equilibrium, it also warrants the continuity equation to be valid
in the form \begin{equation}\label{conteq}\frac{\partial}{\partial
t} n(Q,t) + \frac{\partial}{\partial Q} j(Q,t) =0, \end{equation}
with the current and spatial densities being defined as
\begin{equation}\label{current} j(Q,t) = \int dP \,\frac{P}{M(Q)}
\,f(Q,P,t)\end{equation} and \begin{equation}\label{spadens}
n(Q,t) = \int dP \,f(Q,P,t)\,,\end{equation} respectively. This is
easily verified with the help of (\ref{KramersQP}) exploiting
partial integrations with respect to momentum $P$ (for which no
"surface terms" survive).

Both in Liouville's equation as well as in the transport equation
(\ref{KramersQP}) a term appears which needs to be treated with
special care. It is the one which results from the spatial
derivative of the kinetic energy, and reads
\begin{equation}\label{KramZusatz} \left( \frac{\partial}{\partial
Q} \,\frac{P^{2}}{2 M(Q)} \right) \frac{\partial}{\partial P}
\,f(Q,P,t)\,.
\end{equation} Indeed, this term is absent in the common
derivation of the rate formula (see e.g. \cite{hantalbor}), where
always a constant inertia is assumed to be given. It may be noted
that this term may imply difficulties with "saddle point
approximation", as needed in Kramers' stationary solution.
Moreover, and more important for the present purpose, this term is
also neglected in extracting the transport coefficients within the
linear response approach. There, a locally harmonic approximation
is exploited, which for the sake of simplicity is formulated only
with respect to the coordinate. What that means may be visualized
in the following way. Look at the Liouville part of the transport
equation, in particular at the term which involves the derivative
of the Hamilton function with respect to $Q$. First of all there
is the ordinary force from the potential, which in the expansion
around a $Q_0$ to harmonic order may be written as
\begin{equation}\label{potforce} -\frac{\partial}{\partial Q} V(Q)
\equiv K(Q) \approx  K(Q_0) - C(Q_0)(Q-Q_0)\,,\end{equation} with
the second derivative of the potential defining the local
stiffness $C(Q_0)$. In addition there is the term
\begin{equation}\label{harmfull} \left( \frac{\partial}{\partial
Q} \,\frac{P^{2}}{2 M(Q)} \right) =P^{2}\frac{\partial}{\partial
Q} \left( \frac{1}{2 M(Q)}\right)\,
\end{equation} which is of second order (namely in $P$). In a
consistent treatment one would have to introduce a $P_0$ and
expand all terms locally in collective phase space around the
$(Q_0,P_0)$ to second order both in $Q-Q_0$ {\it and} $P-P_0$ (see
\cite{HOFNGO}). So far this has not been done when extrapolating
transport coefficients from the microscopic linear response
approach.  This does not imply that our inertia may not change
with the coordinate at all. It only means that the terms including
its derivative have been discarded. With respect to the basic
transport equation (\ref{KramersQP}) this approximation may be
phrased as
\begin{equation}\label{appvarinert} \left| \left(
\frac{\partial}{\partial Q} \,\frac{P^{2}}{2 M(Q)} \right)
\frac{\partial}{\partial P} \,f(Q,P,t)\right| \;\ll \;\left|
\frac{\partial}{\partial P} \left( - K(Q)+ \frac{P}{M(Q)}
\,\gamma(Q) \right) f(Q,P,t)\right|\,.
\end{equation}

\subsection{Kramers' decay rate formula extended to variable inertia}\label{decravarin}

For reasons given above, we still want to make use of the
condition (\ref{appvarinert}). Nevertheless, it is necessary to
re-derive an expression for the rate, which still is defined as
\begin{equation}\label{def-rate} R = \frac{j_{b}}{N_{a}}\,.
\end{equation}
To calculate the quantities involved we need a global solution
$f_{\text{glob}}(Q,P)$ of equation (\ref{KramersQP}) which
corresponds to a small but finite, (quasi-)stationary current
across the barrier. This current may be calculated from
(\ref{current}), with $f(Q,P,t)$ replaced by
$f_{\text{glob}}(Q_{b},P)$.  The probability $N_{a}$ of finding
the system inside the well at $Q_{a}$ may be calculated as follows
\begin{equation}\label{numba} N_{a} = \int\limits_{Q_{a} - \Delta}\limits^{Q_{a} +
\Delta} dQ \int dP \,f_{\text{glob}}(Q,P) = \int\limits_{Q_{a} -
\Delta}\limits^{Q_{a} + \Delta} dQ \,n_{\text{glob}}(Q)\,.
\end{equation}
The integration range $2\Delta$ has to be smaller than $Q_{b} -
Q_{a}$ but large enough such that it contains the vast majority of
the ensemble points sitting in the well. An approximation for
$f_{\text{glob}}(Q,P)$ may be constructed by matching together at
some intermediate point $\overline{Q}$ local solutions valid at
the minimum $Q_{a}$ and at the barrier $Q_{b}$. The global
solution $f_{\text{glob}}(Q,P)$ will have an overall normalization
factor which drops out when calculating the rate from
(\ref{def-rate}). The normalization of the local solutions $f(Q
\approx Q_{a}, P)$ and $f(Q \approx Q_{b}, P)$, on the other hand,
might be different. It has to be chosen in appropriate fashion
such that both solutions match properly at $\overline{Q}$.

For a sufficiently high barrier the particles inside the potential
well may be assumed to stay close to a local equilibrium
associated to the temperature $T$. In case that the corresponding
fluctuations $\langle \Delta Q^2\rangle_a^{\text{eq}}$ concentrate
on a region around the minimum, the potential may be replaced by a
harmonic oscillator and the associated local phase-space density
may be approximated by: \begin{equation}\label{fmin}
f_a(Q,P)\equiv f(Q \approx Q_{a}, P) = {\cal N}_{a} \,\exp \left(
-\beta \left( \frac{P^{2}}{2 M_{a}} + V(Q_{a}) + \frac{C_{a}}{2}
\,\left(Q - Q_a\right)^{2} \right) \right)\,.
\end{equation} This is a reasonable estimate up to a $\overline{Q}$
with \begin{equation}\label{condwimin}\left(\overline{Q} -
Q_a\right)^{2} \gtrsim \langle \Delta Q^2\rangle_a^{\text{eq}}
\equiv \frac{T}{C_a}\,.
\end{equation}%
Likewise, approximating the barrier by an inverted oscillator, the
phase-space density may be represented by Kramers' stationary
solution \cite{kram, hantalbor} (neglecting any quantum effects,
see e.g. \cite{hofrep}) \begin{equation}\label{fbarr}
\begin{split}f_b(Q,P)\equiv f(Q \approx Q_{b}, P)  = & {\cal
N}_{b} \,\exp \left( -\beta \left( \frac{P^{2}}{2 M_{b}} +
V(Q_{b}) - \frac{|C_{b}|}{2} \,(Q - Q_{b})^{2} \right) \right)  \\
& \times \int_{-\infty}^{P - A (Q - Q_{b})} du
\,\frac{1}{\sqrt{2\pi\sigma}} \,\exp \left( -\frac{u^{2}}{2\sigma}
\right) \,, \end{split}
\end{equation}
where \begin{equation}\label{defsigA} A =
\frac{|C_{b}|}{\varpi_{b} \left( \sqrt{1 + \eta_{b}^{2}} -
\eta_{b} \right)}
\qquad \text{and} \qquad
\sigma = T M_{b}\Biggl( \frac{1}{\left( \sqrt{1 +
\eta_{b}^{2}} - \eta_{b} \right)^{2}} - 1\Biggr)
\end{equation}

This latter solution must be joined to the one given by
(\ref{fmin}) in some intermediate region, say at $\overline{Q}$.
It would be too much to require this to be possible for {\it any}
$P$, in particular as we are not able to treat the inertia in
continuous fashion, for reasons given above. However, as the rate
is determined by the ratio of two quantities which are averaged
over momentum it turns out sufficient to match only the reduced
$Q$-space densities. Like suggested by (\ref{spadens}), the latter
are obtained by integrating out the momentum $P$ in (\ref{fmin})
and (\ref{fbarr}) to get
\begin{equation}\label{nmin} n_{a}(Q) = {\cal N}_{a} \,\sqrt{2\pi
M_{a}T} \,\exp \left( -\beta \left( V(Q_{a}) + \frac{C_{a}}{2} (Q
- Q_{a})^{2} \right) \right)
\end{equation}
and \begin{equation}\label{nbarr} n_{b}(Q) = {\cal N}_{b}
\,\sqrt{\frac{\pi M_{b} T}{2}} \,\exp \left( -\beta \left(
V(Q_{b}) - \frac{|C_{b}|}{2} (Q - Q_{b})^{2} \right) \right)
\left( 1 + \textrm{erf} \left( \sqrt{\frac{|C_{b}|}{2T}} \,(Q -
Q_{b}) \right) \right)
\end{equation}
To obtain the last expression identities for error functions have
been applied. Still it is not of a form for which a condition like
$ n_a(\overline{Q}) = n_b(\overline{Q})$  would make much sense.
For this we need the additional assumption
\begin{equation}\label{condwidebarr} |C_{b}| \,(\overline{Q} -
Q_{b})^{2} \gg T \,,\end{equation} which renders the
error-function in (\ref{nbarr}) close to unity. Then the two
densities match smoothly at a $\overline{Q}$ satisfying
\begin{equation}\label{conpotqb}V(Q_{a}) + \frac{C_{a}}{2} (\overline{Q} -
Q_{a})^{2} =  V(Q_{b}) - \frac{|C_{b}|}{2} (\overline{Q} -
Q_{b})^{2} \,,
\end{equation} if only the normalization constants are
chosen according to \begin{equation}\label{normab} {\cal N}_{a} =
{\cal N}_{b} \,\sqrt{\frac{M_{b}}{M_{a}}} \,.
\end{equation}

Now we are in the position to calculate the rate from
(\ref{def-rate}). Plugging (\ref{normab}) into (\ref{fmin}) the
number of "particles" at the minimum (\ref{numba}) becomes:
\begin{equation}\label{Na} N_{a} \approx {\cal N}_{b} \,\sqrt{2 \pi M_{b}T}
\,\sqrt{\frac{2 \pi T}{C_{a}}} \,\exp ( -\beta V(Q_{a})) = {\cal
N}_{b} \,2 \pi T \,\sqrt{\frac{M_{b}}{C_{a}}} \,\exp ( -\beta
V(Q_{a}))
\end{equation}
To get this simple expression it was assumed that the $\Delta$,
which in (\ref{numba}) defines the range of integration, is of the
order of or larger than the fluctuation $\langle \Delta
Q^2\rangle_a^{\text{eq}}$ given in (\ref{condwimin}), such that
the Gaussian integral can be calculated for
$\Delta\rightarrow\infty$.  The current at the barrier may be
evaluated from (\ref{fbarr}) and (\ref{current}) with $Q = Q_{b}$.
After a lengthy but straight forward calculation involving
identities for error-integrals, once more, one arrives at
\begin{equation}\label{jb} j_{b} = {\cal N}_{b} \,T
\,\sqrt{\frac{M_{b}}{|C_{b}|}} \,\varpi_{b} \left( \sqrt{1 +
\eta_{b}^{2}} - \eta_{b} \right) \,\exp (-\beta V(Q_{b}))
\end{equation}
From (\ref{jb}) and (\ref{Na})  the decay rate (\ref{def-rate})
turns into \begin{equation}\label{rateKramers}
\begin{split}R_K^{h.v.}  = & \frac{\varpi_{b}}{2\pi}
\,\sqrt{\frac{C_{a}}{|C_{b}|}} \left( \sqrt{1 + \eta_{b}^{2}} -
\eta_{b} \right) \,\exp (-\beta E_{b})
\\  = & \frac{\varpi_{a}}{2\pi} \,\sqrt{\frac{M_{a}}{M_{b}}}
\left( \sqrt{1 + \eta_{b}^{2}} - \eta_{b} \right) \,\exp (-\beta
E_{b})\,,  \end{split}
\end{equation}
confirming the expression (\ref{krahiviin}) used in the text. We
may note again that the result (\ref{rateKramers}) was derived
before in \cite{rumhof} within an extension of  the Perturbed
Static Path Approximation (PSPA).

Finally, we like to come back once more to the conditions
(\ref{condwimin}) and (\ref{condwidebarr}) imposed before. They go
along with the relation (\ref{conpotqb}) for $\overline{Q}$ and
the barrier height $E_b=V(Q_{b})-V(Q_{a})$. The latter must then
satisfy $E_b = \frac{C_{a}}{2} (\overline{Q} - Q_{a})^{2} +
\frac{|C_{b}|}{2} (\overline{Q} - Q_{b})^{2} \gg T $. It may be
useful to visualize these relations with the help of the following
schematic potential:
\begin{equation}\label{potschem} V(Q)=
\begin{cases}
V(Q_{a}) + \frac{C_{a}}{2} (Q - Q_{a})^{2} & \text{for $Q <
\overline{Q}$}, \\[4pt]  E_{b} + V(Q_{a}) - \frac{|C_{b}|}{2} (Q -
Q_{b})^{2} & \text{for $Q > \overline{Q}$}\,.\\
\end{cases}
\end{equation}
Choosing the $\overline{Q}$ according to $\overline{Q} = (C_{a}
Q_{a} + |C_{b}| Q_{b})/(C_{a} + |C_{b}|)$ the two parabolas match
smoothly with a continuous first derivative. Possible errors
related to (\ref{condwimin}) and (\ref{condwidebarr}) may easily
be estimated from elementary properties of the error function.

\subsection{The relation to the transition state result}\label{rate-trans}

In transition state theory one assumes to be given a system that
is totally equilibrized inside the barrier and for which at the
barrier current only flows outward discarding any backflow. Within
its most general version, the fission rate has been estimated by
N. Bohr and J.A. Wheeler through their famous formula
\cite{bohrwheel}. There the equilibrium is the one of a micro
canonical ensemble as represented by the density of states (of the
total system at the minimum and of the intrinsic system at the
barrier). To the extent that the micro canonical ensemble may be
represented by a canonical one, with one and the same temperature
at the minimum and at the barrier, the calculation of the rate can
be done as follows, looking only at the collective degree of
freedom. The outward current (at the barrier) is given by
\begin{equation}\label{curr-tran} j_b^{trans} =
\int_0^\infty \,dP\,\frac{P}{M_b} f_{eq}(Q_b,P) \;\propto\;
\int_{-\infty}^\infty \,dP\,\frac{P}{M_b}
\exp\left(-\beta\frac{P^2}{2M_b}\right)\,\Theta(P)\,.\end{equation}
Comparing with Kramers's stationary solution shown in
(\ref{fbarr}) one realizes \cite{swiat-priv} the only difference
being the replacement of the Theta function $\Theta(P)$ of
(\ref{curr-tran}) by the integral (for $Q=Q_b$) which appears in
the second line of (\ref{fbarr}). Because of the following
representation of the Theta function
\begin{equation}\label{reptheta}
  \Theta(P) = \lim_{\sigma \to 0}\int_{-\infty}^P du
\,\frac{1}{\sqrt{2\pi\sigma}} \,\exp \left( -\frac{u^{2}}{2\sigma}
\right) \,,
\end{equation}
it is seen that (for finite temperature)
\begin{equation}\label{rat-tr-kr}
  R_{trans}=R_K^{h.v.}(\eta=0)\,.
\end{equation}
This follows immediately with the help of the expression given for
$\sigma$ in (\ref{defsigA}).

\subsection{The Smoluchowski limit}\label{smolueq}

Performing in (\ref{rateKramers}) the limit $\eta_{b} \gg 1$, with
the effective damping rate defined by (\ref{defgameta}), one gets:
\begin{equation}\label{rateSmol} \begin{split} R_K^{h.v.}
 \, \longrightarrow & \,
\frac{\varpi_{b}}{2\pi} \,\sqrt{\frac{C_{a}}{|C_{b}|}}
\,\frac{1}{2 \eta_{b}} \,\exp (-\beta E_{b})\,=\,  \frac{1}{2\pi}
\,\frac{1}{\gamma_{b}} \,\sqrt{C_{a} |C_{b}|} \,\exp (-\beta
E_{b})  \\  = & \; R_{\text{ovd}} \,.\end{split}\end{equation}
This expression coincides with the formula (\ref{smolrate})
associated above with the Smoluchowski limit. As a matter of fact,
this result can be obtained directly from the Smoluchowski
equation \begin{equation}\label{SmolQP} \frac{\partial}{\partial
t} n(Q,t) = \frac{\partial}{\partial Q} \left( \frac{1}{\gamma(Q)}
\,\frac{\partial V(Q)}{\partial Q} + \frac{T}{\gamma(Q)}
\,\frac{\partial}{\partial Q} \right) n(Q,t) =
\frac{\partial}{\partial Q} \,j(Q,t) \,.
\end{equation}
The calculation of the decay rate is quite easy in this case.
Indeed, eq.(\ref{SmolQP}) gives an explicit form of the $j(Q,t)$
as a function of the density $n(Q,t)$. Although in our case the
friction coefficient varies with $Q$ the common derivation of the
rate formula (see e.g. \cite{Risken}) may be taken over without
much difficulties.

Notice, please, that the transition from an equation like
(\ref{KramersQP}) to (\ref{SmolQP}) only requires the $\eta(Q)$ to
be large enough at {\it any} $Q$. Such a transition may be
performed also in the case of a variable inertia, at least if
condition (\ref{appvarinert}) is fulfilled. In any case, in the
overdamped limit the inertia has to drop out. We may note in
passing that this transition is in accord with the locally
harmonic approximation in the form discussed in sect.2.2.5 of
\cite{hofrep}. Following the arguments of sect.10.1 and 10.4 of
\cite{Risken} eq.(\ref{SmolQP}) can strictly be derived from
(\ref{KramersQP}) neglecting the term (\ref{KramZusatz}).

\subsection{Strutinsky's derivation of the rate formula}\label{kram-stru}

The essential idea exploited in \cite{strutkram} is written there
below eq.(6). Different to the approach described in
sect.\ref{decravarin}, the number of particles $N_a$ at the
minimum is estimated by multiplying the density $n_b(Q)$ of
Kramers' stationary solution calculated at $Q_a$ by an "effective
length", which in turn is determined by the mean fluctuation of
the oscillator at this $Q_a$ times $\sqrt{2\pi}$, viz by
\begin{equation}\label{equfluct} \sqrt{2\pi}\left(\Delta Q\right)_{\text{eq}}
= \sqrt{\frac{2\pi T}{C_a}}\,.\end{equation} (The additional
factor $\sqrt{2\pi}$ is required to ensure the appropriate measure
needed for the normalization of a Gaussian). In this way one gets
from (\ref{jb}) and (\ref{nbarr})
\begin{equation}\label{ra-Stru-Kra} R_K^{h.v.} =
\frac{\varpi_{b}}{2\pi} \,\sqrt{\frac{C_{a}}{|C_{b}|}} \left(
\sqrt{1 + \eta_{b}^{2}} - \eta_{b} \right) \,\exp (-\beta
E_{b})\,.\end{equation} Here, it was assumed (i) that the
$Q_b-Q_a$ is sufficiently large such that in (\ref{nbarr}) the
error function could be replaced by unity, and (ii) that the
barrier height can be estimated as $E_b \approx (|C_{b}|/2) (Q_a -
Q_{b})^{2}$. The result (\ref{ra-Stru-Kra}) has the same form as
given in (\ref{rateKramers}). As explained earlier, it is
equivalent to (\ref{krahiviin}) or the second line of
(\ref{rateKramers}) which involve the inertias. This latter
expression is identical to the one given in eq.(16) of
\cite{strutkram} if {\em one only interchanges there primed and
unprimed quantities}.
\section{A schematic microscopic model} \label{schemod}

\subsection{The Lorentz-model for intrinsic motion} \label{lomoint}

Let us assume that the nucleonic excitations can be parameterized
by the response function (in this section we set $\hbar=1$)
\begin{equation}\label{resint} \chi(\omega)= - \overline{F^2} \;
\left[\frac{1}{\omega - \Omega + i\Gamma/2}  - \frac{1}{\omega +
\Omega+ i\Gamma/2} \right]\end{equation} Here, the average matrix
element $\overline{F^2}$ of the one body operator $\hat F$, which
acts as the generator of collective motion, measures the overall
strength of the distribution. The states reached by that coupling
are centered at $\Omega$ with an effective bandwidth $\Gamma$
(measured here in units of $~\text{MeV}/\hbar$). For real
frequencies the reactive and dissipative response functions,
$\chi^{\prime}$ and $\chi^{\prime\prime}$, are readily calculated
noticing that they represent real and imaginary parts of
$\chi(\omega)=\chi^\prime(\omega)+i\chi^{\prime\prime}(\omega)$.
The static response is given by
\begin{equation}\label{statres} \chi(0) = \frac{2\Omega\overline{F^2}}
{\Omega^2 + (\Gamma/2)^2} =\chi^\prime(\omega=0) \ .
\end{equation} It is useful to rewrite this (intrinsic) response
function $\chi(\omega)$ in terms of the form of the oscillator
response given in (\ref{defoscres}).
\begin{equation}\label{oscint} \chi(\omega)= \frac{-2
\Omega\overline{F^2}}{\omega^2 + i\Gamma \omega - (\Omega^2 +
(\Gamma/2)^2)}= \frac{-1/M_{\rm int}}{\omega^2 + i\Gamma_{\rm int}
\omega - \varpi^2_{\rm int}}\,.\end{equation} In this way
transport coefficients for intrinsic motion appear
\begin{equation}\label{inttranco} M_{\rm int}=\frac{1}{2\Omega\overline{F^2}
}, \qquad \Gamma_{\rm int}=\Gamma, \qquad \varpi^2_{\rm int}=
\Omega^2 + (\Gamma/2)^2\,. \end{equation}

Next we turn to the collective response. For the $F$-mode it is
given by \cite{hofrep} \begin{equation}\label{collres}
\chi_{\text{coll}}(\omega) = \frac{\chi(\omega)}{
1+k\chi(\omega)}=\frac{1}{\frac{1}{\chi(\omega)}
+k}=\frac{-1/M_F}{\omega^2 + i\Gamma_F \omega - \varpi^2_F}\
,\end{equation} with the inverse coupling constant
\begin{equation}\label{coupcon} -\frac{1}{k} = C(0) + \chi(\omega=0);
\end{equation} and $C(0)$ being the stiffness of the free energy.
The transport coefficients for the collective $F$-mode are:
 \begin{equation}\label{coltranco} M_F=M_{\rm int}, \qquad
 \Gamma_F= \Gamma_{\rm int}, \qquad C_F\equiv
 M_F\varpi^2_F= M_{\rm int}\varpi^2_{\rm int} + k
\end{equation} To get those for the $Q$-mode one needs to multiply these
quantities by $1/k^2$. For slow modes it so turns out that to a
good approximation the $C(0)$ in (\ref{coupcon})  may be neglected
as compared to the $\chi(0)$. This leads to
\begin{equation}\label{inert1m} M=\frac{1}{k^2}M_F\approx
\frac{\bigl(\chi(0)\bigr)^2}{2\Omega\overline{F^2} } =
\frac{2\Omega\overline{F^2} }{\bigl(\Omega^2 +
(\Gamma/2)^2\bigr)^2} = \frac{\chi(0)}{\Omega^2 + (\Gamma/2)^2} =
\frac{1}{2}\, \left.\frac{\partial^2\chi^\prime}{\partial
\omega^2} \right|_{\omega=0} \,.\end{equation} In the last
expression we have made use of (\ref{statres}). Last not least
this has been done because the static response seems to be quite
insensitive to the increase of temperature \cite{kidhofiva}, at
least for not too large $T$. The transformation from the $F$- to
the $Q$- mode leaves ratios between transport coefficients
unchanged. The collective width, the ratio between friction and
inertia, thus becomes $ \Gamma_{\rm kin} = \Gamma$. For friction
this implies
\begin{equation}\label{frict1m} \gamma=\Gamma_{\rm kin} M =
\frac{\Gamma}{\Omega^2 + (\Gamma/2)^2}\,\chi(0) =
\left.\frac{\partial \chi^{\prime\prime}}{\partial
\omega}\right|_{\omega=0} \,,\end{equation} and for the stiffness
one gets the expected result $ C \approx C(0)$, which follows
because the $C_F$ of (\ref{coltranco}) can be written as
\begin{equation}\label{stiffcal}C_F= M_F\varpi^2_F=
\frac{1}{\chi(0)} + k = \frac{C(0)}{\chi(0)\Bigl(C(0) +
\chi(0)\Bigr)}\approx
\frac{C(0)}{\Bigl(\chi(0)\Bigr)^2}.\end{equation} The second
equation follows from (\ref{inttranco}) and (\ref{statres}).

Finally, we should like to note that for the schematic model with
only one mode the inertia always is the one which defines the
value of the energy weighted sum. Likewise, as one may see from
(\ref{frict1m}) for friction and (to lesser extent) from
(\ref{inert1m}) for the inertia, these transport coefficients are
well represented by their "zero frequency limits".

\subsection{Benefits and shortcomings of this
model}\label{benshort}

\subsubsection{Weak damping} \label{weakdam}

For this model weak damping is defined as $ \Omega \gg \Gamma/2$.
In this case static response and inertia turn into the expressions
known from the so called "degenerate model" \cite{bm} $ \chi(0) =
2\overline{F^2}/\Omega$ and $ M= 2\overline{F^2}/\Omega^3$. Notice
that the inertia has the typical structure of the cranking
inertia. For friction one gets $\gamma = 2 \Gamma
\overline{F^2}/\Omega^3 $.

The degenerate model becomes most transparent if it is applied to
the case that nucleons move in oscillator potentials, in
particular if any spin dependent forces are neglected. Then the
intrinsic excitation is given by 
$\hbar\Omega = \Delta N \hbar\Omega_0 \equiv \Delta N \ (41
~\text{MeV}/A^{1/3})$, where $\Delta N$ is the difference in the
major quantum numbers of those states which are coupled through
the multipole operator $F$. Whereas for the quadrupole there is
only one possibility, namely $\Delta N=2$, this is no longer true
for other multipoles, for which more than just one mode are
possible. The same holds true as soon as a spin orbit force is
introduced. Then even for the quadrupole transitions with $\Delta
N= 0$ are possible. It is them which lead to the low frequency
modes we are typically interested in, as they resemble closest the
fission mode. If one still likes to stick to the (degenerate or)
Lorentz model --- which only allows for one mode
---  the effective frequency $\Omega$ will only be a fraction of the
shell spacing parameter $\Omega_0$.

\subsubsection{Strong damping} \label{strongdam}

It is tempting to apply this schematic model also to the extreme
case of very strong damping when $\Gamma$ becomes comparable to or
larger than the frequency $\Omega$ of the typical intrinsic
excitation. Plain confidence into the formula (\ref{frict1m})
would lead to $ \gamma \simeq (4/\Gamma)\chi(0)$. This seems
particularly intriguing if on the one hand the static response
does indeed not change much with $T$, and if, on the other hand,
the $\Gamma$ is associated to the widths the single particle
states, as will be discussed below. According to (\ref{widthspwi})
there might then be some range in which the friction force would
show the typical $1/T^2$ dependence one expects for liquids in the
"collision dominated regime", see also sect.5.3 of \cite{hoivya}.
However, we claim that for finite nuclei the situation is more
complicated. Evidently, the effects of strong collisions are due
to the increasing importance of residual interactions. But the
latter imply other consequences as well, last and not least a
mixing with more complicated states such that with increasing
thermal excitations many particle - many hole states become more
and more important. As has been demonstrated in previous papers,
see e.g. \cite{hoyaje, hofrep} amongst others, this effect implies
that high frequency modes shift to lower frequencies such that the
typical mode at stake in the transport model gets more and more
strength --- implying that finally its inertia is given by the sum
rule limit. Moreover, it has been demonstrated that this feature
goes along with the disappearance of shell effects at $T=T_{\rm
shell}$. This problem is addressed in the text.

\subsubsection{Temperature dependence through collisional damping}
\label{tdepcolda}

Looking back at the intrinsic response function introduced in
(\ref{resint}), one realizes that the only quantity which can be
expected to change sensitively with excitation is the width
$\Gamma$. To get some first orientation we may relate it to the
single particle width. For a Fermi system the latter can be
expected to be of the form \cite{hofrep} \footnote{Different to
the notation used in \cite{hofrep} here energies are measured with
respect to the Fermi surface $\mu$}
\begin{equation}\label{spwidth}
\Gamma_{\rm sp}(\omega,T)= \frac{1}{\Gamma_0}\;\frac{(\omega
-\mu)^2+\pi^2T^2}{1+ \bigl((\omega -\mu)^2+\pi^2T^2\bigr)/c^2}\,,
\end{equation} with the parameters \begin{equation}\label{defgacu}
\frac{1}{\Gamma_0} = 0.03 ~\text{MeV}^{-1} \qquad {\rm and }
\qquad c= 20 ~\text{MeV}\,.\end{equation} For slow collective
motion we may omit the frequency dependence and evaluate this
width at the Fermi surface and put $\omega-\mu=0$ in
(\ref{spwidth}) . Along this approximation we may put
\begin{equation}\label{widthspwi} \Gamma_{\rm kin} \approx
2\Gamma_{\rm sp}(\mu,T) \approx \frac{0.6T^2}{1+ T^2/40}
~\text{MeV} \qquad (T ~{\rm in} ~\text{MeV})\,.
\end{equation} Evidently the correction term in the denominator
only becomes important at temperatures of the order of $T\simeq
6~\text{MeV}$. This is already beyond that value were the other
effects come into play we discussed in sect.\ref{strongdam}. For
this reason the actual $\Gamma(T)$ is changed in the main text.

Finally we may note that our schematic model is not capable of
accounting for pairing. The latter will modify the transport
properties at temperatures below $T\simeq T_{\rm pair}$. This is
discussed in the text.

\vspace*{0.5cm} \noindent {\bf Acknowledgements}

The authors gratefully acknowledge financial support by the
Deutsche Forschungsgemeinschaft, and they are grateful to J.
Ankerhold, B. Back, A. Kelic and M. Thoennessen for enlightening
discussions. Two of us (F.A.I. and S.Y) would like to thank the
Physik Department of the TUM for the hospitality extended to them
during their stay.

\end{document}